\documentstyle[12pt]{article}
\advance \topmargin by -\headheight
\advance \topmargin by -\headsep

\evensidemargin \oddsidemargin
\begin{document}

\topmargin -.6in

%
\def\rf#1{(\ref{eq:#1})}
\def\lab#1{\label{eq:#1}}
\def\nonu{\nonumber}
\def\br{\begin{eqnarray}}
\def\er{\end{eqnarray}}
\def\be{\begin{equation}}
\def\ee{\end{equation}}
\def\foot#1{\footnotemark\footnotetext{#1}}
\def\lb{\lbrack}
\def\rb{\rbrack}
\def\llangle{\left\langle}
\def\rrangle{\right\rangle}
\def\blangle{\Bigl\langle}
\def\brangle{\Bigr\rangle}
\def\llbrack{\left\lbrack}
\def\rrbrack{\right\rbrack}
\def\lcurl{\left\{}
\def\rcurl{\right\}}
\def\({\left(}
\def\){\right)}
\def\v{\vert}
\def\bv{\bigm\vert}
\def\Bgv{\;\Bigg\vert}
\def\bgv{\bigg\vert}
\def\dot3{\cdot\cdot\cdot}
\def\lskip{\vskip\baselineskip\vskip-\parskip\noindent}
\relax

\def\tr{\mathop{\rm tr}}
\def\Tr{\mathop{\rm Tr}}
\def\partder#1#2{{{\partial #1}\over{\partial #2}}}
\def\funcder#1#2{{{\delta #1}\over{\delta #2}}}

\def\a{\alpha}
\def\b{\beta}
\def\d{\delta}
\def\D{\Delta}
\def\eps{\epsilon}
\def\vareps{\varepsilon}
\def\g{\gamma}
\def\G{\Gamma}
\def\grad{\nabla}
\def\h{{1\over 2}}
\def\l{\lambda}
\def\L{\Lambda}
\def\m{\mu}
\def\n{\nu}
\def\o{\over}
\def\om{\omega}
\def\O{\Omega}
\def\p{\phi}
\def\P{\Phi}
\def\pa{\partial}
\def\pr{\prime}
\def\ra{\rightarrow}
\def\s{\sigma}
\def\S{\Sigma}
\def\t{\tau}
\def\th{\theta}
\def\Th{\Theta}
\def\z{\zeta}
\def\ti{\tilde}
\def\wti{\widetilde}

\def\phanta{\phantom{Never Communism}}
\def\phantb{\phantom{Never Islamic Fundamentalism}}
\def\phantc{\phantom{Never Communism and Islamic Fundamentalism}}

\def\lie{{\cal G}}
\def\dlie{{\cal G}^{\ast}}
\def\elie{{\widetilde \lie}}
\def\edlie{{\elie}^{\ast}}
\def\hlie{{\cal H}}
\def\wlie{{\widetilde \lie}}
\def\f#1#2#3 {f^{#1#2}_{#3}}                   

\def\rlx{\relax\leavevmode}
\def\inbar{\vrule height1.5ex width.4pt depth0pt}
\def\IZ{\rlx\hbox{\sf Z\kern-.4em Z}}                 
\def\IR{\rlx\hbox{\rm I\kern-.18em R}}                
\def\IC{\rlx\hbox{\,$\inbar\kern-.3em{\rm C}$}}       
\def\one{\hbox{{1}\kern-.25em\hbox{l}}}
\def\0#1{\relax\ifmmode\mathaccent"7017{#1}%
        \else\accent23#1\relax\fi}
\def\omz{\0 \omega}

\def\ltimes{\mathrel{\raise0.04cm\hbox{${\scriptscriptstyle |\!}$
\hskip-0.175cm}\times}}

\def\mark{\noindent{\bf Remark.}\quad}
\def\prop{\noindent{\bf Proposition.}\quad}
\def\theor{\noindent{\bf Theorem.}\quad}
\def\name{\noindent{\bf Definition.}\quad}
\def\exam{\noindent{\bf Example.}\quad}


\def\Ouc{{\cal O}_{(U_0 ,c)}}           
\def\Gsu{G_{stat}(U_0 ,c)}                   
\def\Gs{G_{stat}}                            
\def\Asu{{\lie}_{stat} (U_0 ,c)}                   
\def\As{{\lie}_{stat}}                             
\def\Suc#1{\Sigma \Bigl( #1 ; (U_0 ,c) \Bigr)}        
\def\suc#1{{\hat \sigma} (#1 ; (U_0 ,c))}
\def\sh{\hat s}                              
\def\ssh#1{{\hat \sigma}^{#1}}
\def\Y#1{Y(#1)}
\def\y{{\hat y}}
\def\yp{y_{+}(g^{-1})}
\def\YT{Y_t (g^{-1})}
\def\yt{Y_t (g)}
\def\W1#1{W \lbrack #1 \rbrack}                
\def\Wuc#1{W \lbrack #1 ; (U_0 ,c)\rbrack}

\def\hd{{\widehat D}}
\def\dt{{\hat d}}
\def\Gpr{G^{\pr}}
\def\GT{\tilde \Gamma}
\def\GTy{\funcder {\GT} {y (t)}}
\def\GTz#1{\funcder {\GT} {y_{#1} (t)}}
\def\Ly#1{{\hat L}^{#1}_t (y)}
\def\Ry#1{R^{#1}_t (y)}
\def\LA{{\cal L}^A}

\def\Tu#1{{\widetilde \Theta}^{#1}}
\def\Td#1{{\widetilde \Theta}_{#1}}
\def\Z{\widetilde Z}
\def\T#1{{\hat {\cal T}}(#1)}
\def\dNz{\delta^{(N)} (z_1 - z_2 )}
\def\dNth{\delta^{(N)} (\th_1 - \th_2 )}
\def\dN#1#2{\delta^{(N)} ({#1} - {#2})}
\def\DN{{\llbrack D {\widetilde \Theta} \rrbrack}^2_N}
\def\Du#1{{\widetilde D}^{#1}}
\def\Dd#1{{\widetilde D}_{#1}}

\def\Tor{{\wti {\rm SDiff}}\, (T^2 )}
\def\Lh#1{{\hat {\cal L}}({#1})}
\def\M{{\cal M}}
\def\dM{{\cal M}^{\ast}}
\def\Mc{{\cal M}(R^1 \times S^1 )}
\def\dMc{{\cal M}^{\ast}(R^1 \times S^1 )}
\def\st1{\stackrel{\ast}{,}}

\def\Winf{{\bf W_\infty}}
\def\Win1{{\bf W_{1+\infty}}}
\def\DO{DOP (S^1 )}                           
\def\DA{{\cal DOP} (S^1 )}                    
\def\eDA{{\widetilde {\cal DOP}} (S^1 )}       
\def\dDA{{\cal DOP}^{\ast} (S^1 )}                  
\def\edDA{{\widetilde {\cal DOP}}^{\ast} (S^1 )}    
\def\DOP#1{{DOP (S^1 )}_{\geq{#1}}}             
\def\DAP#1{{{\cal DOP} (S^1 )}_{\geq{#1}}}                    
\def\eDAP#1{{{\widetilde {\cal DOP}} (S^1 )}_{\geq{#1}}}
\def\dDAP#1{{{\cal DOP}^{\ast} (S^1 )}_{\geq{#1}}}     
\def\edDAP#1{{{\widetilde {\cal DOP}}^{\ast} (S^1 )}_{\geq{#1}}}
\def\PsDO{\Psi{\cal DO} (S^1 )}           
\def\Rm#1#2{r(\vec{#1},\vec{#2})}          
\def\sto{\stackrel{\circ}{,}}              
\def\sta{\, ,\,}
\def\xx{(\xi , x)}
\def\yy{(\eta , y)}
\def\xxt{(\xi , x ; t )}
\def\intres{\int dx\, {\rm Res}_\xi \; }
\def\intrest{\int dt\, dx\, {\rm Res}_\xi \;}
\def\Res{{\rm Res}_\xi}
\def\pexx{e^{\pa_x \pa_\xi}}
\def\mexx{e^{-\pa_x \pa_\xi}}
\def\SLinf{SL (\infty ; \IR )}             
\def\slinf{sl (\infty ; \IR )}               
\def\sumlm{\sum_{l=1}^{\infty} \sum_{\v m\v \leq l}}
\def\WDO#1{W_{DOP (S^1 )} \lb #1\rb}               

\def\A{\cal A}
\def\AA{\widetilde {\cal A}}
\def\ruc#1{r_{#1} (U_0 ,c)}                      

\newcommand{\nit}{\noindent}
\newcommand{\ct}[1]{\cite{#1}}
\newcommand{\bi}[1]{\bibitem{#1}}

\begin{titlepage}

\noindent\null\hfill {{\sl BGU-92 / 11 / July - PH}}

\noindent\null\hfill {{\sl hep-th/9207048}}

\vskip .6in

\begin{center}
{\large {\bf $\, {\bf W_\infty}\,$ Gravity - a Geometric
         Approach}}
\end{center}

\normalsize
\vskip .4in

\begin{center}
{E. Nissimov$^{\,1}$ and
S. Pacheva\footnotemark
\footnotetext{On leave from: Institute of Nuclear Research and
Nuclear Energy, Boul. Trakia 72, BG-1784 Sofia, Bulgaria.}}

\par \vskip .1in \noindent
Department of Physics \\
Ben-Gurion University of the Negev \\
Box 653, 84105 Beer Sheva, Israel \\
\par \vskip .3in

\end{center}

\begin{center}
Submitted to : {\em Theoretical and Mathematical Physics}
\end{center}

\vskip .2in

\begin{center}
{\em In Memoriam}
\end{center}

\begin{center}
{\large{\bf M. C. Polivanov}}
\end{center}

\vskip .3in

\begin{center}
{\large {\bf ABSTRACT}}\\
\end{center}
\par \vskip .3in \noindent

A brief review is given of an adaptation of the coadjoint orbit
method appropriate for study of models with infinite-dimensional
symmetry groups. It is illustrated on several examples, including
derivation of the WZNW action of induced $D=2\,$ $(N,0)\,$ supergravity.
As a main application, we present the geometric action
on a generic coadjoint orbit of the deformed group of area preserving
diffeomorphisms. This action is precisely the anomalous effective
WZNW action of $D=2 \,$ matter fields coupled to chiral $\Winf$ gravity
background. Similar actions are given which produce the {\em KP}
hierarchy as on-shell equations of motion.

\end{titlepage}

\indent
{\sl
For the international scientific community Professor M.C. Polivanov was a
highly esteemed theoretical physicist in a world-renown institution
where the foundations of numerous branches of modern theoretical and
mathematical physics ranging from axiomatic quantum field theory
\ct{Poliv} to soliton theory and quantum groups \ct{Fadd} took shape.
But for the frequent foreign visitors to Steklov Mathematical
Institute he was much more. We shall always remember with deep
admiration his profound and all embracing erudition, invariably accompanied
by the noble human warmth of a genuine Russian aristocrat.}
\lskip \noindent
{\large{\bf 1. Introduction}}
\lskip \indent
In the last few years a lot of attention was devoted to the
infinite-dimensional Lie algebra $\,\Winf$ and its generalizations
$\,{\bf W_{1+\infty}}$ etc. \ct{Pope1,Pope2,Bakas}.
These algebras are nontrivial ``large $N$"
limits of the associative, but {\em non}-Lie, conformal $\,{\bf W_N}$
algebras \ct{Za85}.
{}From a purely algebraic point of view, $\Winf\,$ is isomorphic to the
centrally extended Lie algebra $\eDA\,$ of differential operators of
arbitrary order on the circle \ct{dop,RadVays92}.
Geometrically, it is a nontrivial deformation of the Lie algebra
$\, \bf w_\infty\,$ of area preserving diffeomorphisms on cylinder.

$\bf W_{(1)+\infty}\,$ algebras arise in various problems of two-dimensional
physics : self-dual gravity \ct{selfdual}, first Hamiltonian structure of
integrable Kadomtsev-Petviashvili
({\em KP}) hierarchy \ct{Wata84,KP}, string field actions in the collective
field theory approach \ct{AvJev91}, stringy
black holes \ct{Wblack}, conformal affine Toda theories
\ct{AFGZ}. One of the most remarkable manifestations of $\,\Winf$-type
algebras is the recent discovery of a subalgebra of their ``classical" limit
$\, {\bf w_\infty}\,$ in $c=1$
string theory as symmetry algebra of the special discrete states
\ct{KlePo91} or as the algebra of infinitesimal deformations of the
ground ring \ct{Wit91}.

A characteristic feature of all conformal field theory models,
including those based on $\bf W_{(1)+\infty}$ algebras, is that
their dynamics is entirely described in terms of an underlying
infinite-dimensional Noether symmetry algebra containing the Virasoro
algebra as a subalgebra. Thus, to study them it is natural to invoke
\ct{AlFaSh89} the general theory of group coadjoint orbits \ct{KK}
extending it to the infinite-dimensional case.

In section 2 below we briefly outline a general group coadjoint orbit
formalism appropriate to study geometric actions and their symmetries
for arbitrary infinite-dimensional groups with central extensions.
In section 3 solution of Ward identities for quantum effective
(WZNW) actions
in terms of geometric actions on group coadjoint orbits is discussed.
As nontrivial examples, we present the explicit form of the geometric
actions associated with $D=2\,$ $(N,0)\,$ super-Virasoro group (for any
$\, 0\leq N \geq 4\,$ ), which are the anomalous effective actions of
induced $D=2\,$ $(N,0)\,$ supergravity. In section 4 the group coadjoint
orbit formalism is applied to derive the WZNW effective action of
induced $\Winf$ gravity. Also, Lagrangian actions are given which yield
the {\em KP} hierarchy as equations of motion.
\lskip
{\large{\bf 2. Brief Review of Coadjoint Orbits and Geometric Actions}}
\lskip
\noindent{\bf 2.1 Basic Ingredients}

Let us consider arbitrary (infinite-dimensional)
group $G$ with a Lie algebra $\lie$ and its dual space
$\dlie$ . The adjoint and coadjoint actions of $G$ and
$\lie$ on $\lie$ and $\dlie$ are given by
$\,Ad(g)\, X = g\, X\, g^{-1} \; ,\; ad(X) \, Y = \left\lb X ,
\, Y \right\rb\;\,$ and $\,\; \llangle Ad^{*}(g) U \v X \rrangle =
\llangle U \v Ad(g^{-1}) X \rrangle \; ,\; \llangle ad^{*}(X)U \v
Y \rrangle = \;$ $\,- \llangle U \v \lb X , Y \rb \rrangle $
 Here $g \in G$ and $X ,\, Y \in \lie \; ,\; U \in
\dlie$ are arbitrary elements, whereas $\langle \cdot \v
\cdot \rangle\,$ indicates the natural bilinear form
``pairing" $\lie$ and $\dlie$ .

For theoretical physics applications, the primary interest lies
in infinite-dimensional Lie algebras with a
central extension $\elie = \lie \oplus \IR$ of $\lie \;$ and,
correspondingly, an extension $\edlie = \dlie \oplus \IR$
of the dual space $\dlie$. The central extension is given
by a linear operator $ \sh : \lie \longrightarrow \dlie$
satisfying
\be
\sh (\lb X , Y \rb ) =
ad^{\ast}(X) \sh (Y) - ad^{\ast}(Y) \sh (X)  \lab{jacob}
\ee
which defines a nontrivial two-cocycle on the Lie algebra $\lie$ :
\be
\om (X ,Y ) \equiv - \l \llangle \sh (X )\bv Y \rrangle
\qquad \qquad \forall \, X ,Y \in \lie
\lab{acocycle}
\ee
where $\l$ is a model-dependent normalization constant.
The Jacobi identity \rf{jacob}  can be integrated ($Y
\longrightarrow g= \exp Y $) to get a unique nontrivial
$\dlie$-valued group one-cocycle $S(g)$ in terms of the Lie-algebra
cocycle operator $\sh$ (provided $H^1 (G) = \emptyset \,
,\, {\rm dim} H^2 (G) = 1 \,$; see \ct{Ki82} ) :
\be
ad^{*} (X ) S(g)  = Ad^{*}(g)\, \sh \( Ad(g^{-1}) X \)
- \sh (X )  \qquad\qquad \forall \, X \in \lie
\lab{kir}
\ee
satisfying the relations :
\be
\sh (X ) = {d \o d \t} S( e^{\t X}) \Bgv_{\t =0} \;\;\;
,\;\;\; S(g_1 g_2 ) = S(g_1 ) + Ad^{*}(g_1 )S(g_2 )  \lab{cocycle}
\ee
Now, we can easily generalize the adjoint and coadjoint
actions of $G$ and $\lie$ to the case with a central
extension (acting on elements $(X ,n),(Y ,m) \in \elie\,$ and
$(U,c) \in \edlie $; see e.g. \ct{ANPZ}):
\br
{\wti A}d (g) \,(X , n) &=& \left( Ad (g) X \;, \; n + \lambda
\llangle S (g^{-1}) \, \v \, X \rrangle \right)
\label{eq:eadj}  \\
{\wti ad} (X ,n ) \; (Y ,m )
&\equiv&  \llbrack (X ,n ) \, , \, (Y ,m ) \rrbrack =
 \Bigl( \;\lb X , Y \rb \; ,\; -
\l \llangle \sh (X ) \v Y \rrangle  \Bigr)
\lab{ealgadj}
\er
\be
{\wti Ad}^{\ast} (g) \, (U , c) = \( Ad^{\ast}(g) U + c \l \, S (g)
 \;, \; c \) \;\;\; ,\;\;\;
{\wti ad}^{*} (X ,n) (U,c) = \left( ad^{*} (X ) \,U + \,c \,\l
\, \sh (X )\; , \; 0 \;\right)  \lab{ecoadj}
\ee
Also, the bilinear form $\langle \cdot \v \cdot \rangle$
on $\dlie \otimes \lie$ can be extended to a bilinear form
on $\edlie \otimes \elie$ as :
\be
\llangle (U,c) \bv (X ,n) \rrangle =
\llangle U \bv X \rrangle + c\, n     \lab{ebilin}
\ee
The physical
interpretation of the $\lie$-cocycle $\sh $ is that of ``anomaly" of the
Lie algebra (i.e., existence of a c-number term in the commutator
\rf{ealgadj} ), whereas
the group cocycle $S(g)$ is the integrated ``anomaly", i.e. the ``anomaly"
for finite group transformations (see eqs.\rf{eadj} and \rf{cocycle} ).

Another basic geometric object is the fundamental
$\lie$-valued Maurer-Cartan one-form $Y(g)$ on $G$
satisfying $\,
d \, Y(g) = \h \llbrack \, Y(g) \, , \, Y(g) \, \rrbrack\,$.
Here and in what follows $d\,$ denotes exterior derivative. $Y(g)\,$
is related to the group one-cocycle $S(g)$ through the
equation :
\be
d \, S (g ) = ad^{\ast} (Y(g))\, S(g) + \sh (Y(g))
\label{eq:comvary}
\ee
and possesses group one-cocycle property similar to that
of $S(g)$ \rf{cocycle} :
\be
Y(g_1 g_2 ) = Y(g_1 ) + Ad(g_1 ) Y(g_2 ) \lab{ycocycle}
\ee
The group- and algebra-cocycles
$S(g)$ and $\sh (X)$ can be generalized to
include trivial (co-boundary) parts
($\,(U_0 ,c)$ being an arbitrary point in the
extended dual space $\edlie \,$) :
\br
\S (g) & \equiv& \Suc{g}  = c\l S(g) + Ad^{\ast}(g)U_0 - U_0
\lab{Sigmab}  \\
\ssh{} (X )& \equiv& \suc{X} = ad^{*} (X ) U_0 + c\l \,\sh (X )
= {d \o d \t} \S (e^{ \t X}) \Bgv_{\t =0}
\lab{sigmab}
\er
The generalized
cocycles \rf{Sigmab} and \rf{sigmab} satisfy the same relations
as \rf{cocycle}, \rf{comvary} and \rf{kir}.
\lskip
\noindent{\bf 2.2 Coadjoint Orbits}

The coadjoint orbit of $G$, passing through the point
$(U_0 ,c)$ of the dual space $\edlie$, is defined as (cf.
\rf{ecoadj} ) :
\be
\Ouc \equiv \biggl\{ \Bigl( U (g),c \Bigr) \in \edlie \; ;\;
U(g) =  U_0 + \S (g) = Ad^{*}(g) U_0 + c\l S(g) \biggr\} \lab{orbit}
\ee
The orbit \rf{orbit} is a right coset
$\Ouc \simeq G / {\Gs}$ where $\Gs\;$ is the stationary subgroup of the
point $(U_0 ,c)$ w.r.t. the coadjoint action \rf{ecoadj} :
\be
\Gs = \biggl\{ k \in G \; ;\; \S(k) \equiv
c\l S(k) + Ad^{\ast}(k)U_0 - U_0 = 0 \biggr\} \lab{gstat}
\ee
The Lie algebra corresponding to $\Gs$ is :
\be
\As \equiv \biggl\{ X_0 \in \lie \; ;\; {\hat \s} (X_0 )
\equiv ad^{*} (X_0 ) U_0 + c\l \,\sh (X_0 ) = 0 \biggr\}
\lab{algstat}
\ee
The physical meaning of $\As\,$ is that of maximal ``anomaly-free"
subalgebra of $\lie$ , i.e. the maximal subalgebra on which the central
extension vanishes (cf.\rf{eadj}-\rf{ecoadj}).

Now, using the basic geometric objects from sect. 2.1,
we can express the Kirillov-Kostant symplectic form $\O_{KK}\,$ \ct{KK}
on $\Ouc$ for any infinite-dimensional (centrally extended) group $G$
in a simple compact form \ct{ANPZ} . Namely,
introducing the centrally extended objects :
\be
{\wti \S}(g) \equiv (\S (g),c) \in \edlie \;\;\; ,\;\;\;
{\wti Y}(g) \equiv \Bigl( Y(g), m_Y (g) \Bigr) \in \elie
\lab{centr}
\ee
\be
d{\wti \S}(g) = {\wti ad}^{\ast} \Bigl({\wti Y}(g)\Bigr)
{\wti \S}(g) \;\;\; ,\;\;\; d{\wti Y}(g) = \h \llbrack
{\wti Y}(g)\, ,\, {\wti Y}(g) \rrbrack  \lab{tilde}
\ee
we obtain (using \rf{ebilin} and \rf{comvary}) :
\be
\O_{KK} = -d \biggl( \llangle {\wti \S}(g) \bv {\wti Y}(g)
\rrangle \biggr) = -\h \llangle d {\wti \S}(g) \bv {\wti
Y}(g) \rrangle                  \lab{KK}
\ee
\noindent{\bf 2.3 Geometric Actions and Symmetries}

The geometric action on a coadjoint orbit $\Ouc$ of
arbitrary infinite-dimensional (centrally extended) group
$G$ can now be written down compactly as \ct{ANPZ,kovhid} :
\be
{\wti W} \lb g\rb = \int d^{-1} \O_{KK} - \int dt \,H  =
- \int \llangle {\wti \S}(g) \bv {\wti Y}(g) \rrangle
- \int dt \, H\left\lb {\wti \S}(g) \right\rb           \lab{action}
\ee
In more detail, introducing the explicit expressions
\rf{centr}, \rf{tilde}, \rf{Sigmab} and \rf{sigmab}, the ``kinetic" part
of \rf{action} reads :
\be
\W1{g} = \int \llangle U_0 \bv Y(g^{-1}) \rrangle -
c\l \int \biggl\lb \Bigl\langle S(g) \bv Y(g) \Bigr\rangle - \h
d^{-1} \Bigl( \Bigl\langle \sh (Y(g)) \bv Y(g) \Bigr\rangle\Bigr)
\biggr\rb  \lab{gaction}
\ee
The integral in \rf{action}, \rf{gaction} is over
one-dimensional curve on the phase space $\Ouc$ with a
``time-evolution" parameter $t$ . Along the curve the
exterior derivative becomes $\, d = dt\, \pa_t$ and the
projection of the one-form $Y(g)$ is : $Y(g)=dt \,\yt$.
Note also the presence of the multi-valued (in general)
$\, d^{-1}\,$ term \ct{nov} on the r.h.s. of \rf{gaction}.


The group cocycle properties of $S(g)$ and $Y(g)$ (
eqs.\rf{cocycle} and \rf{ycocycle} ) imply the following
fundamental group composition law \ct{kovhid}
(with $\S (g)$ as in \rf{Sigmab}) :
\be
\W1{g_1 g_2} = \W1{g_1} + \W1{g_2} + \int \llangle \S
(g_2 ) \bv Y(g_1^{-1}) \rrangle      \lab{pw}
\ee
Eq.\rf{pw} is a
generalization of the well-known Polyakov-Wiegmann composition
law \ct{PW} in WZNW models to geometric actions on
coadjoint orbits of arbitrary groups with central
extensions.

The whole symmetry structure of the geometric action \rf{gaction} is
encoded in eq.\rf{pw}. Indeed, considering first
arbitrary right group translations $
g\longrightarrow g \exp (\t {\large \vareps} )\,$, ${\large \vareps}
\in \lie \,$ , we get from \rf{pw} :
\be
{d\o {d\t}} \W1{g\,\exp (\t {\large \vareps} )} \Bgv_{\t =0} \equiv
\d^R_{\large \vareps} \W1{g} = \int \llangle \ssh{}({\large \vareps} )
\bv Y(g^{-1}) \rrangle
= - \int \llangle \ssh{} (Y(g^{-1})) \bv {\large \vareps} \rrangle
\lab{right}
\ee
Recalling \rf{algstat} one finds ``gauge" invariance of
$\W1{g}$ under right group translations from the
stationary (``anomaly-free") subgroup $\Gs\,$ \rf{gstat} of the orbit $\Ouc\,
$ \rf{orbit} : $\d^R_{{\large\vareps}_0} \W1{g} = 0\,$ for arbitrary
time-dependent ${\large\vareps}_0 (t) \in \As\,$ \rf{algstat}.
This fact demonstrates the geometric meaning of
``hidden" local symmetries \ct{D2grav} in models with
arbitrary infinite-dimensional Noether symmetry groups.

Next, for arbitrary left group translations $g\longrightarrow \exp
(\t {\large \vareps})\, g\,$ one obtains, using \rf{pw}, the Noether theorem :
\be
{d\o {d\t}} \W1{\exp (\t {\large \vareps}) \, g} \Bgv_{\t =0} \equiv
\d^L_{\large \vareps}
\W1{g} = - \int \llangle \S (g) \bv d{\large \vareps}  \rrangle
\lab{left}
\ee
i.e., $\S (g)\,$ \rf{Sigmab} is a Noether conserved
current when the Hamiltonian is absent in \rf{action}
: $\pa_t \S (g) = 0$ . More generally, one gets the equations of motion :
\be
\pa_t \S (g) + ad^{\ast}\left(\funcder{H}{\S (g)}\right) \S (g)
+ {\hat \s} \left(\funcder{H}{\S (g)}\right) = 0
\lab{motion}
\ee
using notations \rf{Sigmab},\rf{sigmab}.
\lskip
{\large{\bf 3.Quantum Effective Actions from Geometry}}
\lskip
{\bf 3.1 Geometric Actions and Ward Identities}
\lskip \indent
Let us consider the Legendre transform of the geometric action $\W1{g}$
\rf{gaction} on a coadjoint orbit of $G$ :
\be
\G\lb g\rb \equiv \W1{g} - \int \Bigl\langle \S (g) \bv
\funcder{\W1{g}}{\S (g)} \Bigr\rangle = \W1{g} + \int
\llangle \S (g) \bv Y(g) \rrangle = - \W1{g^{-1}}
\lab{legendr}
\ee
One can easily show that, as a functional of $\; y \equiv \yt\,$,
\rf{legendr} satisfies the functional equation :
\be
\pa_t \funcder{\G}{y} - ad^{\ast}(y )
\funcder{\G}{y} - {\hat \s} (y ) = 0  \lab{ward}
\ee
As shown in \ct{kovhid,shorty}, eq.\rf{ward} coincides with
the renormalized Ward identity for the following functional integral :
\be
\exp i\G \lb y\rb = \int {\cal D}\Phi \exp i \Bigl\{ W_0 \lb \Phi\rb
+ \int \llangle J(\Phi ) \bv y \rrangle \Bigr\}
\lab{funcint}
\ee
The notations in \rf{funcint} are as follows. $W_0 \lb \Phi\rb\,$ is a
classical action of ``matter" fields $\, \Phi_a\,$ possessing an
infinite-dimensional Noether symmetry group $\, G_0\,$
with generators $\, \{ T^I\}\,$ .
The index $I\,$ is a short-hand notation for $\,
I=\Bigl(\; (x_1 ,...,x_p );A\; \Bigr) \,$ ,
including in general both continuous parameters
$(x_1 ,...,x_p ) \;$ (e.g., in the case of $p$-brane models) as well as
discrete indices $A$ (as in the case of Kac-Moody groups).
The Noether group $G_0\,$ is
in general a contraction (``classical" limit) of $\, G\,$ .
The corresponding Noether conserved currents $J^I (\Phi ) = T_I^{\ast}
\, J^I (\Phi ) \,$, which belong to the dual space $\dlie_0 \simeq
\dlie\, $, span Poisson bracket algebra of the form:
\be
\{ J^I (\Phi ), J^K (\Phi ) \}_{PB} = - \omz^{IK} - \omz_L^{IK} J^L (\Phi )
\lab{pbj}
\ee
where $\,\omz_L^{IK}\,$ and $\,\omz^{IK}\,$ denote the structure
constants and the (possible) central extension of the ``classical"
Noether Lie algebra $\,\lie_0\,$ w.r.t. the basis $\{ T^I \}\,$ .

Thus, the Legendre transform of the $G$ co-orbit action \rf{legendr}
is the exact solution for the quantum effective action $\,\G \lb y\rb
\,$ \rf{funcint} no matter what is the specific form of the classical
``matter" action $\, W_0 \lb \Phi\rb\,$ provided the Noether symmetries
of the latter form a group $G_0\,$ coinsiding with, or being contraction
of, $\, G\,$ .
\lskip
{\bf 3.2 Examples of Geometric Actions}
\lskip
{\bf 3.2.1 Kac-Moody Groups}

The Kac-Moody group elements $g \simeq g(x)$ are smooth
mappings $S^1 \longrightarrow G_0\,$, where $G_0$ is a
finite-dimensional Lie group with generators $\{ T^A \}$
 The explicit form of \rf{eadj}-\rf{ecoadj} reads in this case :
\br
Ad(g)X = g(x)X (x) g^{-1}(x) \;\;\; &,&\;\;\;
ad(X_1 ) X_2 = \lb X_1 (x)\, ,\, X_2 (x) \rb
\;\;\; , \;\;\; X_{1,2} (x) = X_{1,2}^A (x) T_A \nonu
\\
Ad^{\ast}(g) U = g(x) U(x) g^{-1}(x) \;\;\; &,& \;\;\;
ad^{\ast}(X )U = \lb X (x) \, ,\, U(x) \rb \;\;\; ,
\;\;\; U(x) =U_A (x) T^A  \nonu
\er
\be
\sh (X ) = \pa_x X (x) \;\;\; ,\;\;\; S(g) = \pa_x g
(x) \, g^{-1}(x) \;\;\; ,\;\;\; Y(g) = dg(x)\, g^{-1}(x)
\lab{KM}
\ee
Plugging \rf{KM} into \rf{gaction} one obtains the
well-known WZNW action for $G_0$-valued chiral fields
coupled to an external ``potential" $U_0 (x)$.

\lskip
{\bf 3.2.2 Virasoro Group}

The Virasoro group elements $g \simeq F(x)$ are smooth
diffeomorphisms of the circle
$S^1 \;$. Group multiplication is given by composition of
diffeomorphisms in inverse order : $g_1 \cdot \;g_2 = F_2 \circ F_1
\; (x) = F_2\Bigl( F_1 (x)\Bigr)\;$. Eqs.\rf{eadj}-\rf{ecoadj} have now the
following explicit form :
\br
Ad(F) X = {\Bigl( \pa_x F \Bigr)}^{-1} X \( F(x)\) \;\;\; &,&
Ad^{\ast} (F) U = {\Bigl( \pa_x F \Bigr)}^2 U \( F(x)\)
\nonu \\
ad (X )Y \equiv \lb X ,Y \rb = X \pa_x Y - (\pa_x X )
Y
\;\;\; &,&\;\;\; ad^{\ast}(X ) U = X \pa_x U + 2 (\pa_x X) U
\nonumber
\er
\be
\sh (X ) = \pa_x^3 X \;\;\; , \;\;\;
 S(F) = {{\pa_x^3 F}\o {\pa_x F}} - {3\o 2}
{\({{\pa_x^2 F}\o {\pa_x F}}\)}^2 \;\;\; ,\;\;\;
Y(F) = {{dF}\o {\pa_x F}} \lab{coadvir}
\ee
Here $S(F)$ is the well-known Schwarzian.
Plugging \rf{coadvir} into the general expressions
\rf{gaction} and \rf{pw} one reproduces the well-known
Polyakov $D=2\,$ induced gravity action (coupled to an external
stress-tensor $U_0 (x)$ ) :
\be
\W1{F} = \int dt dx \llbrack - U_0 (F(t ,x)) \,\pa_x F\,
\pa_t F + {c\o {48\pi}} {{\pa_t F}\o {\pa_x F}}
\left( {{\pa_x^3 F}\o {\pa_x F}} - 2 {{(\pa_x^2 F)^2}
\o {(\pa_x F)^2}} \right) \rrbrack   \lab{polya}
\ee
and its group composition law \ct{D2grav,AlSh89,Po90}.
\lskip
{\bf 3.2.3 $\bf (N,0)\,$ $\bf D=2\,$ Super-Virasoro Group ($
\bf N \leq 4$)}

Here it is appropriate to use the manifestly
$(N,0)$ supersymmetric formalism. The points of the $(N,0)$
superspace are labeled as $\;(t,z),\;\; z\equiv(x,\th^i ),\; i=1,.., N$ .
The group elements are given by superconformal diffeomorphisms :
\be
z\equiv(x,\th^j)\;\longrightarrow\;
\Z\equiv \Bigl( F(x,\th^j ),\Tu{i}(x,\th^j )\Bigr)      \lab{sconf}
\ee
obeying the superconformal constraints :
\be
D^j F -i\Tu{k}D^j\Td{k} = 0 \;\;\; ,\;\;\;
D^j \Tu{l} D^k \Td{l} - \d^{jk}\DN = 0 \;\;\; ,\;\;\;
\DN \equiv {1\o N}D^m\Tu{n}D_m\Td{n}
\lab{sconstr}
\ee
Here the following superspace
notations are used :
\be
D^i = \partder{}{\th_i} + i \th^i \pa_x
\;\; ,\;\; D^N \equiv {1\o {N!}} \eps_{i_1
\cdot\cdot\cdot i_N} D^{i_1} \cdot\cdot\cdot D^{i_N}  \nonu
\ee
The $(N,0)$ supersymmetric analogues of \rf{coadvir} read:
\be
Ad(\Z )X = {\Bigl( \DN \Bigr)}^{-1} X (\Z (z)) \;\;\; ,\;\;\;
Ad^{\ast}(\Z )U = {\Bigl( \DN \Bigr)}^{2-{N\o 2}} U (\Z (z)) \nonu
\ee
\be
\lb X ,Y \rb = X \pa_x Y - (\pa_x X)
Y - {i\o 2} D_k X D^k Y \;\;\; ,\;\;
ad^{\ast}(X)U = X \pa_x U + (2-{N\o 2}) (\pa_x X )
U- {i\o 2}D_k X D^k U   \nonu
\ee
\be
\sh_N (X )=i^{N(N-2)} D^N \pa_x^{3-N} X \;\;\;
,\;\;\; Y_N (\Z ) = \Bigl( dF + i \Tu{j} d\Td{j} \Bigr)
{\left(\DN \right)}^{-1}            \lab{coadsvir}
\ee
The associated $\dlie$-valued group one-cocycles $S_N(\Z)$ coincide with the
well-known \ct{Scho88} $(N,0)$ super-Schwarzians.
Inserting the latter and \rf{coadsvir} into \rf{gaction}
one obtains the WZNW action of induced $(N,0)\,$ $D=2\,$ supergravity
(i.e., the $(N,0)$ supersymmetric generalization
of the Polyakov $D=2$ gravity action \rf{polya} for any $N \leq 4$)
\ct{ANPZ,ny} :
\br
W_N \lb \Z \rb = \int dt\;(dz) \biggl\lb \pa_t \Bigl( \ln
\DN \Bigr) D^N \pa_x^{1-N} \Bigl( \DN \Bigr) -  \biggr.  \nonu  \\
\biggl. U_0 (\Z ) {\left(\DN \right)}^{1- {N\o 2}}
\Bigl( \pa_t F + i \Tu{j} \pa_t \Td{j} \Bigr)  \biggr\rb    \lab{Naction}
\er
\lskip
{\large{\bf 4. Applications to Induced $\Winf$-Gravity and $KP$ Hierarchy}}
\lskip
{\bf 4.1 Deformations of Algebras of Area-Preserving
Diffeomorphisms}

Let us now concentrate on the algebras $\,\bf w_{\infty}\,$ of area
preserving diffeomorphisms on two-dimensional surfaces.
As shown in \ct{Feig}, the family of possible deformations $\,\Winf (q)\,$
of the initial ``classical" $\, {\bf w_\infty}\,$ depends on
a single parameter $\, q$ and, for each fixed value of $\, q$,
$\,\Winf (q)$ possesses an one-dimensional cohomology with values in
$\IR$ . In particular, for $q=1$ one finds that  $\,\Winf (1)
\simeq \eDA\,$ - the centrally extended algebra of differential operators on
the circle, which was recently studied in refs.\ct{dop}.
The equivalence of $\,\eDA\,$ to the original definition of $\,\Winf (1)
$ \ct{Pope1,Bakas} was explicitly demonstrated in \ct{BaKheKir91}.


More precisely, let us consider the following class of infinite-dimensional
Lie algebras $\, \lie = \DAP{M} \; , \, M=0,1,2,\cdot\cdot\cdot \,$,
of symbols  of differential operators \foot{Let us
recall \ct{trev} the correspondence between (pseudo)differential operators
and symbols :
$\, X \xx = \sum_k \xi^k X_k (x) \longleftrightarrow {\hat X} = \sum_k X_k (x)
\pa_x^k $.} on the circle $S^1$ \ct{RadVays92} :
\be
\DAP{M} =\Bigl\{\, X\equiv X \xx = \sum_{k \geq M}
\xi^k X_k (x)\,\Bigr\}      \lab{dapM}
\ee
$\DAP{1}\,$ is isomorphic to the ``ordinary" $\Winf\,$ algebra, whereas
$\DAP{0}\,$ is isomorphic to $\Win1\,$ algebra.

For any pair $ X,Y \in \lie = \DAP{M} \,$ the Lie commutator is
given in terms of the associative (and  {\em non}commutative) symbol product
denoted henceforth by a cirle $\, \circ\;$ \foot{The coefficients in the
$\xi$-expansion of the symbol product $X\circ Y = \sum_{k\geq M} \left( X
\circ Y \right)_k (x) \,\xi^k $ are given by infinite series. In order
to secure convergence, one might consider rescaling of the differential
operator symbol $\xi$ by a small parameter $\, h\,$ so that (taking e.g.
$M=0$) : $\left( X \circ Y \right)_k (x) = \sum_0^{\infty} h^n X_n (x)
\; \left\{ \sum_0^{{\rm min} (k,n)} h^{k-m} {n\choose m} \pa_x^{n-m} Y_{k-m}
(x) \,\right\}$ .} :
\be
\lb X \sta Y \rb  \equiv X \circ Y - Y \circ X  \;\;\; ;\;\;\;
X \circ Y \equiv X\xx \, \exp \left( \overleftarrow{\pa_\xi}
\overrightarrow{\pa_x} \right) \, Y\xx   \lab{1}
\ee
The dual space $\, \dlie = \dDAP{M} \,$, is defined by factoring the
space $\PsDO = \Bigl\{ U\xx = \sum_{k=1}^{\infty} \xi^{-k}
U_k (x) \Bigr\}$ of all purely pseudodifferential symbols \ct{trev} on $S^1$
modulo the space of ``zero" pseudodifferential symbols $\, U_{\leq M}\xx
\,$  (cf. ref.\ct{NPV92}) :
\br
\dlie = \Bigl\{ U_{\ast}\, ;\;  U_{\ast} \xx = U\xx - U_{\leq M}\xx
\; {\rm for}\; \forall U \in \Psi {\cal DO} \Bigr\}  \lab{3} \\
U_{\leq M}\xx =
\pexx \left( \,\sum_{k=1}^{M} \xi^{-k} \sum_{l=1}^{k} {{k-1}\choose{l-1}}
\pa_x^{k-l} U_l (x) \,\right)   \lab{3a}
\er
with the natural bilinear form :
\be
\llangle U \v X \rrangle \equiv \intres U \circ X = \intres
\left( \, \mexx \, U\xx \, \right)\; X\xx    \lab{2}
\ee
In particular, according to
\rf{2} any ``zero" pseudodifferential symbol of the form \rf{3a}
is ``orthogonal" to any differential symbol $X \in \DAP{M}$ ,
i.e. $\llangle  U_{\leq M} \v X \rrangle = 0 \,$.
Having the bilinear form \rf{2} one can define the coadjoint action of
$\DAP{M}$ on $\dDAP{M}$ via :
\be
\Bigl( ad^{\ast}(X)\, U\,\Bigr) \xx \equiv \lb X \sta U \rb_{\ast} =
\lb X \sta U \rb_{-} - \lb X \sta U \rb_{\leq M} \phanta  \lab{4}
\ee
Here and in what follows, the subscript $\, (-)\,$ indicates taking the part of
the symbol containing all negative powers in the $\xi$-expansion, whereas the
subscript $\,\ast\,$ indicates projecting of the symbol on the dual space
\rf{3} .

The central extension in $\elie \equiv \eDAP{M} = \DAP{M} \oplus \IR\,$ is
given by the two-cocycle $\, \omega (X,Y) =
- {1\o {4\pi}} \llangle \sh (X) \v Y \rrangle\,$, where the cocycle
operator $\, \sh : \lie \longrightarrow \dlie\, $ explicitly reads \ct{dop}
(cf. also \ct{NPV92}) :
\be
\sh (X) = \lb X \sta \ln \xi \rb_{\ast}    \lab{6}
\ee

Let us now consider the Lie group $G = \DOP{M}\,$ defined as exponentiation
of the Lie algebra $\lie =\DAP{M}\,$  \rf{dapM} :
\be
G=\left\{ g\xx = {\rm Exp} X\xx \equiv \sum_{N=0}^{\infty} {1\o {N!}}
{\underbrace{X\xx \circ \cdot\cdot\cdot
\circ X\xx }}_{{\rm N}\;{\rm times}}\;\right\}   \lab{7}
\ee
and the group multiplication is just the symbol product
$\, g \circ h $ . The adjoint and coadjoint action of
$G =\DOP{M}$ on the Lie algebra $\DAP{M}$ and its dual space $\dDAP{M}$,
respectively, is given as :
\be
 Ad (g) X = g \circ X \circ g^{-1} \;\;\;\; ;\;\;\;\;
 Ad^{\ast} (g) U  = \Bigl( g \circ X \circ g^{-1}
\Bigr)_{\ast}   \lab{9}
\ee
\lskip
{\bf 4.2 WZNW Action for Induced $\bf \Winf\,$ Gravity}

After these preliminaries it is easy to write down the explicit expressions
of the two
fundamental objects $\, S(g) \,$  \rf{cocycle}  (the nontrivial
$\dlie$-valued one-cocycle on the group $G$, or the ``anomaly" for finite
group transformations) and  the Maurer-Cartan form $\, Y (g) \,$
interelated through eq.\rf{comvary}, which enter
the construction of the geometric action on a coadjoint orbit of $G =
\DOP{M}$ :
\be
S (g) = - {\Bigl( \, \lb \ln \xi \sta g \xx \rb \circ g^{-1}\xx
\Bigr)}_{\ast}  \;\;\; ,\;\;\;  Y (g) = dg \xx \circ g^{-1}\xx    \lab{14}
\ee
Plugging \rf{14} and \rf{6} in the general formula \rf{gaction}, one obtains
the
co-orbit geometric action (the explicit dependence
of symbols on $\xxt$ will in general be suppressed below ) :
\br
W_{\DOP{M}} \lb g\rb = - \intrest U_0 \circ g^{-1} \circ \pa_t g  +
\phantb \nonu   \\
{c\o {4\pi}} \int \intres \Biggl( \lb \ln \xi \sta g  \rb \circ
g^{-1} \circ \pa_t g  \circ g^{-1}
-\h d^{-1} \Biggl\{ \left\lb \ln \xi \sta dg \circ
g^{-1}\right\rb \wedge \left( dg \circ g^{-1} \right) \Biggr\}\,\Biggr)
\lab{20}
\er
According to the discussion in Section 3,  the Legendre transform
$\G \lb g\rb = - W\lb g^{-1} \rb\,$ of \rf{20} is precisely the
WZNW anomalous effective action of induced $\Winf$ or $\Win1$
gravity for $M=1,0\,$, respectively.
The physical meaning of the first term on the r.h.s. of \rf{20} is that of
coupling of the chiral $\bf W_{(1)+\infty}$ Wess-Zumino field
$\, g=g\xxt\,$ to a chiral $\bf W_{(1)+\infty}$ gravity ``background".
{}From the general formula \rf{pw}
we get the following fundamental group composition law for $\bf W_{(1)+\infty}$
gravity action :
\be
W \lb g\circ h\rb = W \lb g\rb + W \lb h\rb + \intrest \biggl\{\Bigl( \,
h\circ U_0 \circ h^{-1}
- {c\o {4\pi}}\lb \ln \xi \sta h \rb \circ h^{-1} \,\Bigr) \circ g^{-1} \circ
\pa_t g \; \biggr\}    \lab{21}
\ee
The action \rf{20} implies the Poisson brackets :
\be
\Bigl\{ \, S\lb g\rb \xx\; ,\, S\lb g\rb \yy\, \Bigr\}_{PB} =
- {4\pi \o c}
\biggl\lb\, S\lb g\rb \xx + \ln \xi \; \sta \; \d_{DOP} (y,\eta ;x,\xi ) \,
\biggr\rb_{\ast}   \lab{pb}
\ee
where the symbol commutator on the r.h.s. is the projected one as in
\rf{4},
and $\d_{DOP} (\cdot\, ;\cdot ) \in \dDAP{M} \otimes \DAP{M} \,$ denotes the
kernel of the $\,\d$-function on the space of differential operator symbols :
\be
\d_{DOP} (x,\xi ; y,\eta) = \pexx \left(\, \sum_{k=M}^{\infty} \xi^{-(k+1)}\,
\eta^k\, \d (x-y)\, \right)  \lab{23}
\ee
Eq.\rf{pb} is a compact expression of the Poisson-bracket realization of
$\DAP{M}\,$, in particular, for $\bf W_{(1)+\infty} \simeq \DAP{0,1}\,$
 The component fields $S_r (x)\,$ in the $\xi$-expansion of the
pseudodifferential symbol $\, S\lb g\rb \xx = \sum_{r\geq M+1} S_r (x)
\, \xi^{-r}\,$ turn out to be quasi-primary conformal fields of spin $\,
r \geq M$ . The genuine primary fields ${\cal W}_r (x)\,$ are obtained
from $S_r (x)\,$ by adding derivatives of the lower spin fields
$S_q (x) \;, \, q \leq r-1\,$ .

At this point it would be instructive to explicitate formulas \rf{6}
and \rf{14} when the elements of $G=\DOP{1}$ and
$\lie =\DAP{1} \simeq \Winf \,$ are restricted to the Virasoro subgroup
(subalgebra, respectively) :
\br
X\xx = \xi\, \omega (x) \;\;\;\longleftrightarrow \;\;\;
 \omega (x) \pa_x \;\in {\cal V}ir   \phanta \nonu  \\
g\xx = {\rm Exp}\, (\,\xi\,\omega (x)\, ) \;\;\; \longleftrightarrow\;\;
\; F(x) \equiv \exp ( \omega (x) \pa_x )\, x \;\in {\rm Diff\,}(S^1 )  \lab{16}
\er
Substituting \rf{16} into \rf{6} and \rf{14}, one obtains (cf.
eqs.\rf{coadvir} ) :
\br
Y(g) \Bgv_{g\xx = {\rm Exp}\, (\,\xi\,\omega (x)\, )} = \xi {{dF (x)}\o
{\pa_x F(x)}}    \;\;\; ;\;\;\;
\sh (X) = \lb \xi \omega (x) \sta \ln \xi \rb_{\ast} = - {1\o 6} \xi^{-2}
\;\pa_x^3 \omega (x)  + \cdot\cdot\cdot   \nonu  \\
S(g) \Bgv_{g\xx = {\rm Exp}\, (\,\xi\,\omega (x)\, )} = - {1\o 6} \xi^{-2}
\; \left({{\pa_x^3 F}\o {\pa_x F}} - {3\o 2}
{\({{\pa_x^2 F}\o {\pa_x F}}\)}^2 \right) + \cdot\cdot\cdot
\phanta  \lab{17}
\er
The dots in \rf{17} indicate higher order terms $\, O(\xi^k )$, $k\geq
3$, which do not contribute in bilinear forms with elements of $\,{\cal
V}ir\,$ \rf{16}.  Similar formulas (replacing the factors $1/6\,$  in
\rf{17} by $1/3$ ) are obtained for the embedding of
the Virasoro algebra in $\DAP{0} \simeq \Win1\,$ .
Now, inserting \rf{17} into  eq.\rf{20} (for $M=0,1$ ) one reproduces
\rf{polya}, i.e. Polyakov's action of induced $D=2$ gravity
\ct{D2grav,AlSh89}.
In particular, one finds that the $\DAP{M}$ Maurer-Cartan gauge field
$\; Y_t (g^{-1}) = - g^{-1} \circ \pa_t g \,$
is a generalization of Polyakov's $D=2\,$ gravity gauge field
$\, h_{++} = \pa_t F^{-1}/ \pa_x F^{-1}\, $ where $\, F^{-1}\,$ denotes
the inverse Virasoro group element.

We refer to \ct{NPV92} for a further analysis of the symmetry
properties of the $\Winf\,$ gravity action \rf{20},
namely, the ``hidden" Kac-Moody symmetry which turns out to be a
specific differential operator symbol realization of $\SLinf\,$,
and the associated classical Sugawara construction.
\lskip
{\bf 4.3 Action for {\em {\bf KP}} Hierarchy}

Let us now consider the geometric action of the {\em centerless}
$\DAP{0} \simeq \Win1$ (i.e., put $c=0$ in eq.\rf{20})
 with a nontrivial Hamiltonian added (cf. \rf{action}) :
\be
{\wti W}\lb g\rb = - \intrest \left(\, U^{(0)} \circ g^{-1} \circ \pa_t g
\, \right)  -  {1\o {N+1}} \intrest \Bigl( \,\xi + \left( g\circ
U^{(0)} \circ g^{-1}\right)_{-}\,\Bigr)^{N+1}  \lab{kpaction}
\ee
where $\, U^{(0)} \in \PsDO$ is a fixed pseudodifferential symbol
and the exponent in the last term indicates $(N+1)$-th power
w.r.t. symbol product. One can
easily show that \rf{kpaction} yields Hamiltonian equation
of motion precisely coinsiding with the $N$-th equation of the
$KP$ hierarchy \ct{KPlax} of integrable evolution equations.

Indeed, introducing the notation $\, U(g) \equiv \left( g\circ
U^{(0)} \circ g^{-1}\right)_{-}\,$,
 the equation of motion associated with \rf{kpaction} reads
according to \rf{motion} :
\be
\pa_t U(g) + \left\lb \, \left\{\, \Bigl( \xi + U(g) \Bigr)^N
 \right\}_{+}\, ,\, U(g)\, \right\rb_{-} = 0  \lab{kpmotion}
\ee
where the subscript $(+)\,$ denotes taking the purely differential
part (the non-negative $\xi$-power expansion). Using the simple property $\;
\Bigl\lb \, \left\{ {\cal F}\bigl(\xi + U\bigr)\right\}_{+}\, ,\,
\xi + U \;\Bigr\rb_{+} = 0\;$, where $U\,$ is arbitrary purely
pseudodifferential symbol and ${\cal F}\,$ is arbitrary (analytic)
function, one can immediately rewrite eq.\rf{kpmotion} in the form :
\be
\pa_t \Bigl( \xi + U(g) \Bigr)
+ \left\lb \, \left\{\, \Bigl( \xi + U(g) \Bigr)^N
 \right\}_{+}\, ,\, \xi + U(g)\, \right\rb  = 0  \lab{kplax}
\ee
Of course, eq.\rf{kplax} is nothing but the $N$-th equation of the $KP$
hierarchy in the standard Lax form \ct{KPlax}, where $L\equiv \xi + U(g)
= \xi + \sum_{k\geq 1} U_k \lb g\rb (x) \xi^{-k}\,$ is the symbol of the
corresponding pseudodifferential Lax operator. The identification of
\rf{kplax} as hamiltonian equations on a coadjoint orbit of the
centerless $\Win1\,$ was first pointed out in ref.\ct{Wata84}.

Having an action \rf{kpaction} might help quantization of the {\em KP}
model.

\lskip \indent
\lskip \indent
{\bf Acknowledgements.} Most of material presented above was developed
in collaboration with Henrik Aratyn and Igor Vaysburd to whom we extend
our most sincere appreciations. Also, fruitful discussions with Aharon
Davidson and Miriam Cohen are gratefully acknowledged.

\end{document}